\documentclass[aps,prb,twocolumn,groupedaddress,draft,showpacs,intlimits,amsmath,amssymb,floats]{revtex4}
\newcommand{\bk}{\textbf{k}}
\newcommand{\bq}{\textbf{q}}

\usepackage[final]{graphicx}

\begin{document}


\title{Evidence for forward scattering and coupling to acoustic phonon modes in high-T$_c$ 
cuprate superconductors}

\author{S. Johnston$^{1,4}$}
\author{I. M. Vishik$^{1,2,3}$}
\author{W. S. Lee$^{1,2}$}
\author{F. Schmitt$^{1,2}$}
\author{S. Uchida$^{5}$}
\author{K. Fujita$^{6}$}
\author{S. Ishida$^{5}$}
\author{N. Nagaosa$^{7,8}$}
\author{Z. X. Shen$^{1,2,3}$}
\author{T. P. Devereaux$^{1,2}$}
\affiliation{$^1$Stanford Institute for Materials and Energy Science, 
SLAC National Accelerator Laboratory and Stanford University, Stanford, CA 94305, USA}
\affiliation{$^2$Geballe Laboratory for Advanced Materials, Stanford University, Stanford, CA 94305, USA}
\affiliation{$^3$Department of Physics and Applied Physics, Stanford University, Stanford, CA 94305, USA}
\affiliation{$^4$IFW Dresden, P.O. Box 27 01 16, D-01171 Dresden, Germany}
\affiliation{$^5$Department of Physics, Graduate School of Science, University of Tokyo, Bunkyo-Ku, Tokyo 113-0033, Japan}
\affiliation{$^{6}$Laboratory for Atomic and Solid State Physics, Department of Physics, Cornell University, Ithaca, New York 14853, USA}
\affiliation{$^7$Department of Applied Physics, University of Tokyo, Bunkyo-ku, Tokyo 113-8656, Japan}
\affiliation{$^8$Cross-Correlated Materials Research Group (CMRG) and Correlated Electron Research Group 
(CERG), RIKEN-ASI, Wako 351-0198, Japan}
\begin{abstract}
Recent laser angle-resolved photoemission spectroscopy studies  
have established the presence of a new kink in the low-energy nodal dispersion of 
Bi$_2$Sr$_2$CaCu$_2$O$_{8+\delta}$ (Bi-2212). 
The energy scale ($\sim 8-15$ meV) of this kink appears below 
the maximum of the superconducting gap $\Delta_0$.  
Therefore it is difficult to interpret this feature in terms of the usual coupling to a sharp dispersionless mode.
In this paper we examine electron-phonon coupling to the in-plane 
acoustic phonon branch arising 
from the modulation of the screened Coulomb potential. We demonstrate that 
such a coupling has a strong forward scattering peak, and as a consequence, a 
kink occurs in the dispersion at an energy scale shifted by the local gap $\Delta(\bk)$.   
In addition, considerations for the reduction of screening with underdoping  
naturally explains the observed doping dependence of the low-energy kink.  
These results point to a strong coupling to the acoustic branch which is 
peaked in the forward scattering direction and has important  
implications for transport and pairing in the high-T$_c$ cuprates.
\end{abstract}

\date{\today}
\maketitle

Understanding what controls the superconducting transition temperature T$_c$ in the 
cuprates remains one of the major questions in condensed matter physics.  
The large variation in T$_c$ across cuprate families remains unexplained 
despite the near similarity of the CuO$_2$ plane and accompanying strong correlation 
physics that well describes the parent phases of these systems.  
Strong electron correlations are of course important in all the transition metal 
oxides, but the cuprates are special in the sense that they are the most metallic system 
among them. As a result, the cuprates are at the verge of localization/delocalization 
and/or ionicity/covalency. This subtle balance or dichotomy may be key to the 
origin of the robust superconductivity, which requires both strong coupling and 
quantum coherence. 

This battleground has recently been highlighted in angle-resolved photoemission 
(ARPES), where polaronic lattice effects demonstrably alter the spectral function 
from a Lorentzian to Gaussian lineshape\cite{KylePolaron1,KylePolaron2} in undoped systems. The presence of dispersion "kinks" in doped cuprates have been (controversially)
interpreted as coupling to lattice phonons, with stronger couplings than implied 
by density functional calculations.\cite{Giustino, Heid, Sangiovanni} Finally, with 
the advancements of laser- based ARPES, a new energy scale has been observed in 
the nodal region of Bi$_2$Sr$_2$CaCu$_2$O$_{8+\delta}$ (Bi-2212) which manifests 
as a subtle kink in the band dispersion at an energy  8-15 meV.
\cite{Plumb, ZhangPRL2008, RameauPRB2009,Anzai, VishikPRL2010} 

In this paper, we focus on the latter kink in underdoped Bi-2212 (T$_c$ = 55K, UD55, p $\sim$ 0.088) and present data and 
analysis which suggests that the interplay of Coulomb interactions and lattice 
effects as a consequence of ionicity/covalency dichotomy plays an important 
role in the physics of a doped Mott insulator. 
The experimental setup is identical to Ref. \onlinecite{VishikPRL2010}.  The energy 
resolution was 3 meV and no deconvolution methods have been applied to the data.\cite{RameauPRB2009}  
Whereas previous studies focused on the low-energy kink at the node, we have 
done experiments away from the node where the gap is nonzero.  Since the opening of 
a gap will affect the momentum distribution curve (MDC)   
dispersion close to the gap energy, we chose a doping where the low-energy 
kink is very strong (Ref. \onlinecite{VishikPRL2010}) so the two  
features can be easily distinguished.  However, we emphasize that the 
results presented here hold for other dopings in the underdoped regime.

\begin{figure}[t]
 \includegraphics[width=\columnwidth]{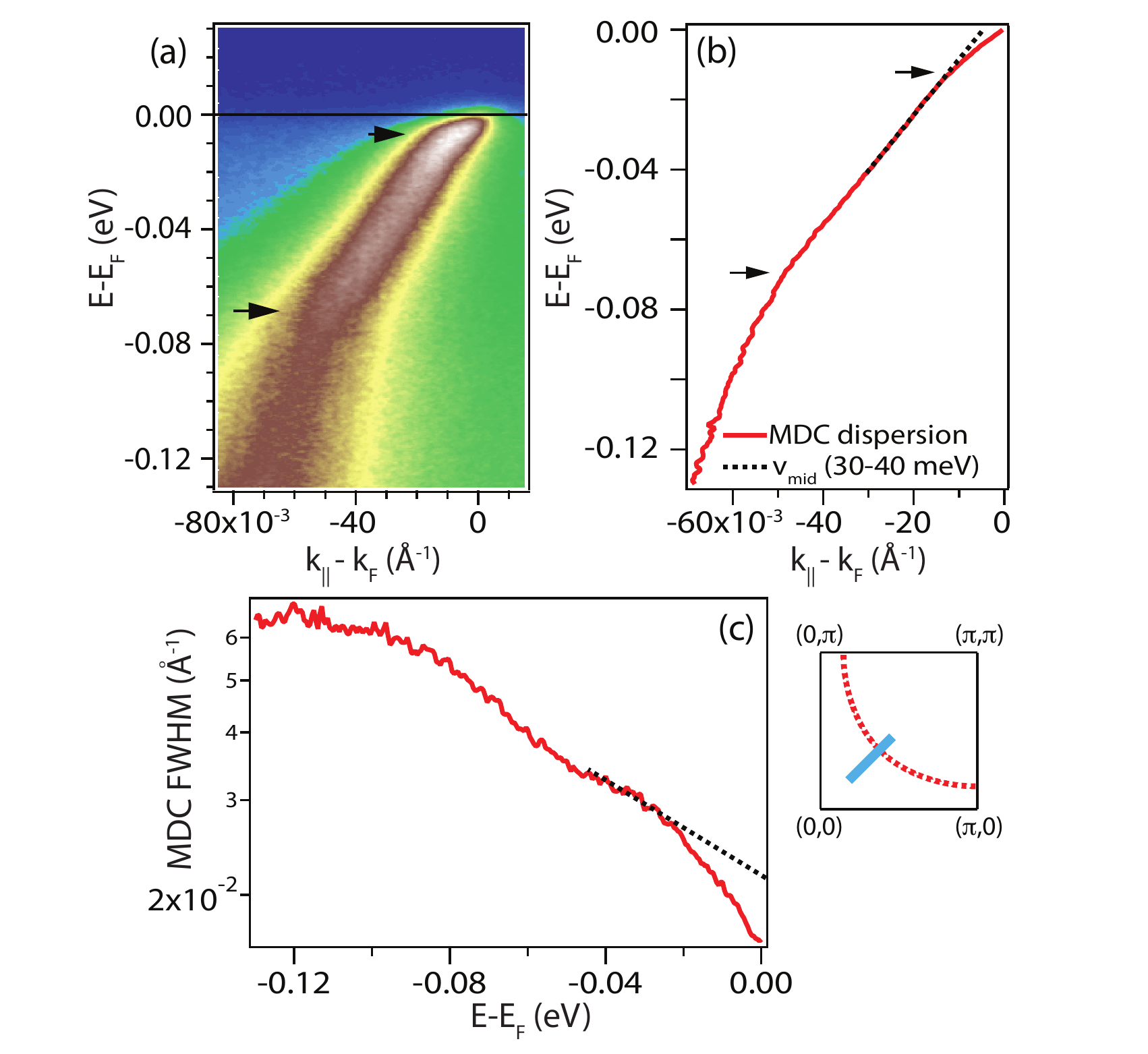}
 \vskip -0.25cm
 \caption{\label{Fig:NodalExp}(Color online) Nodal low-energy kink in Bi-2212 UD55. (a) Color image plot of raw data.  The 70 meV and low-energy kinks are marked by arrows  
 and both can be seen by eye in the raw data via a bend in the dispersion and an 
 accompanying change in spectral intensity. 
 (b) Band dispersion derived by fitting the momentum distribution 
 curve (MDCs, intensity as a function of momentum at fixed energy) at each 
 energy in (a) to a Lorentzian.  The low-energy kink is defined by the deviation 
 of the dispersion from $v_{mid}$, the velocity fit between 30-40 meV (dotted line). 
 (c) MDC FWHM shows a more rapid decrease close to $E_f$ as a consequence of 
 the low-energy kink.}
\end{figure}

The main experimental observations for the low-energy kink along the nodal direction  
are summarized in Fig. \ref{Fig:NodalExp}.  As evident in the raw data (Fig. \ref{Fig:NodalExp}a), 
the kink manifests as a bend in the dispersion and an accompanying change in 
spectral intensity at an energy $\sim 8$-$15$ meV in UD55 Bi-2212. This behavior is also 
evident in the MDC derived dispersion, which deviates from the linear dispersion at 
the same energy scale (Figs. \ref{Fig:NodalExp}b).  
For UD55, the slope of the dispersion at $E_f$ is considerably smaller than the slope of 
the dispersion between 30-40 meV ($v_{mid}$) and the low-energy kink is unmistakable. 
This observation has already been discussed in the context of the universal nodal 
Fermi velocity,\cite{VishikPRL2010} suggesting the importance of this new energy scale 
in the low-energy nodal physics of the high-T$_c$ cuprates.  
A corresponding signature is also seen in the MDC FWHM, which shows a 
more rapid decrease for energies below the low-energy kink.   Additionally, these features 
have been observed over a wide doping range in underdoped Bi-2212, strengthening with 
underdoping, suggesting that it may be a ubiquitous aspect of nodal physics.\cite{VishikPRL2010}

The 70 meV kink has been interpreted in terms of coupling to a sharp 
bosonic mode of energy $\Omega$ which is of either a 
lattice \cite{LanzaraNature2001, JohnstonACMP2010, tpdPRL2004, LeePRB2007, Anzai} or 
magnetic\cite{NormanPRL1997, JohnsonPRL2001, BorisenkoPRL2006, DahmNaturePhysics2009} origin. 
By analogy, it has been suggested that the low-energy kink can be interpreted as coupling to 
an optical phonon.\cite{RameauPRB2009}
In Bi-2212 there are known infrared active optical modes at 97 (12) and 117 cm$^{-1}$ 
(14.5 meV) 
involving $c$-axis motion of the Cu,Ca,Sr and Bi ions\cite{KovalevaPRB2004} 
as well as a Raman active $c$-axis optical branch between 58 and 65 cm$^{-1}$ 
(7.2 and 8.1 meV) with the latter recently invoked to explain the low-energy kink.\cite{RameauPRB2009} 
However, coupling to these optical modes is likely to be 
electrostatic in nature and therefore not sharply peaked in momentum space. 
Such a coupling would would produce a kink at 
$\Omega+\Delta_0$.\cite{Bickers,tpdPRL2004, NormanPRL1997, LeePRB2007} 
However, it is possible to have a non-gap-shifted feature in the 
self-energy near the node if the coupling is strongly peaked in the forward 
scattering direction.\cite{Kulic}  
Such is the case for the in-plane acoustic mode, which we consider here.    

The generic form for the el-ph coupling Hamiltonian is given by 
\begin{equation}
H_{el-ph} = \frac{1}{\sqrt{N}} \sum_{\bk,\bq,\sigma,\nu} |g_\mu(\bk,\bq)|^2 d_{\bk-\bq,\sigma}^\dagger d_{\bk,\sigma} 
(b^\dagger_{\bq,\nu} + b_{-\bq,\nu})
\end{equation}
where $d^\dagger_{\bk,\sigma}$ ($d_{\bk,\sigma}$) creates (annihilates) an electron 
in the antibonding $pd-\sigma^*$ band with momentum $\bk$ and spin $\sigma$ and 
$b^\dagger_{\bq,\nu}$ ($b_{\bq,\nu}$) creates (annihilates) a phonon quanta of momentum $\bq$ in 
branch $\nu$. For the in-plane acoustic branch,
the el-ph interaction arises via a deformation-type coupling 
of the periodic lattice potential.  In this case, momentum-dependent el-ph coupling 
constant $g(\bk,\bq)$ is only a function of $\bq$ and is given by\cite{Mahan}
\begin{equation}\label{Eq:Coupling}
g(\bq) = \frac{1}{V_{cell}}\sqrt{\frac{\hbar}{2M\Omega(\bq)}}\hat{e}_\bq\cdot\bq\frac{V(\bq)}{\epsilon(\bq)}
\end{equation}
with $\Omega(\bq)$ the phonon dispersion, 
$\epsilon(\bq) = 1 + q_{TF}^2/q^2$ the Thomas-Fermi dielectric function, 
$q_{TF}$ the Thomas-Fermi wavevector, $V_{cell} = a^2c$ the unit cell 
volume, $M$ the copper + oxygen ion mass, $\hat{e}_\bq$ the 
phonon polarization vector and $V(\bq) = 4\pi e^2/q^2\epsilon$ the Coulomb 
potential with $\epsilon$ the static dielectric constant.

Apart from the dispersion of $\Omega(\bq)$,  
Eq. (\ref{Eq:Coupling}) has a $\bq$-dependence governed by $\propto \hat{e}_\bq\cdot\bq/(q^2 + q^2_{TF})$ 
which is strongly peaked for $q \sim q_{TF}$ but with $g(\bq)\rightarrow 0$ for 
$\bq \rightarrow 0$.   
In the underdoped cuprates the Thomas-Fermi wavevector is small and 
$|g(\bq)|^2$ is sharply peaked for a subset of small $\bq$.\cite{JohnstonPRB2010}
As a result, the el-ph self-energy is dominated by contributions from scattering 
process from states $E_\bk$ to nearby states $E_{\bk+\bq}$, with  
$E^2(\bk) = \xi^2(\bk)+\Delta^2(\bk)$, where $\xi(\bk)=\epsilon(\bk)-\mu$ is the 
electron dispersion in the normal state measured relative to the Fermi level.\cite{EschrigPRL2000}   
The nodal self-energy is thus determined from scattering to nearby states with 
a small superconducting gap and thus the peak in the self-energy is 
not shifted by the full gap but rather an average of the gap near the node.   

To demonstrate this we now calculate the spectral function 
for coupling to the acoustic phonon branch with 
$\Omega(\bq) = \Omega_0\sqrt{\sin^2(q_xa/2) + \sin^2(q_ya/2)}/\sqrt{2}$ 
and $\Omega_0 = 15$ meV.\cite{Shell}  We also include coupling 
to the 55 and 36 meV in- and out-of-phase $c$-axis polarized modes 
(the so-called $A_{1g}$ and $B_{1g}$ branches, respectively), 
as well as the $70$ meV Cu-O bond-stretching modes 
in order to also capture the well-known 70 meV kink and renormalizations at 
higher binding energy.  
Since our focus is on the features at low energy, we treat the optical modes 
as dispersionless with a momentum-independent coupling. 
The superconducting gap is modelled with a pure $d$-wave form 
$\Delta(\bk) = \Delta_0[\cos(k_xa)-\cos(k_ya)]/2$, where 
$\Delta_0 = 37$ meV is the maximum value at $(0,\pi)$. 
We take the dielectric constant to be set by a large in-plane value with 
$\epsilon = 30\epsilon_0$ (in this context we regard $\epsilon$ as a free 
parameter set to obtain overall agreement with experiment) and take $q_{TF} = 0.5\pi/a$. Finally, 
we take the effective lattice constants per plane to 
be $a = b = 3.8$ and $c = 7.65$ $\AA$, appropriate for Bi-2212.\cite{KovalevaPRB2004} 
The self-energy $\Sigma(\bk,\omega)$ and spectral function $A(\bk,\omega)$ are 
calculated within Migdal-Eliashberg 
theory, which is the same formalism used in prior work to understand the higher energy 
renormalizations\cite{tpdPRL2004,LeePRB2007,JohnstonPRB2010}.  
In order to introduce increased broadening in the spectral function 
with underdoping, we evaluate the self-energies while adding a finite imaginary 
part ($\delta = 7$ meV) to the bare electron Green's function.  

\begin{figure}[tr]
 \includegraphics[width=1.0\columnwidth]{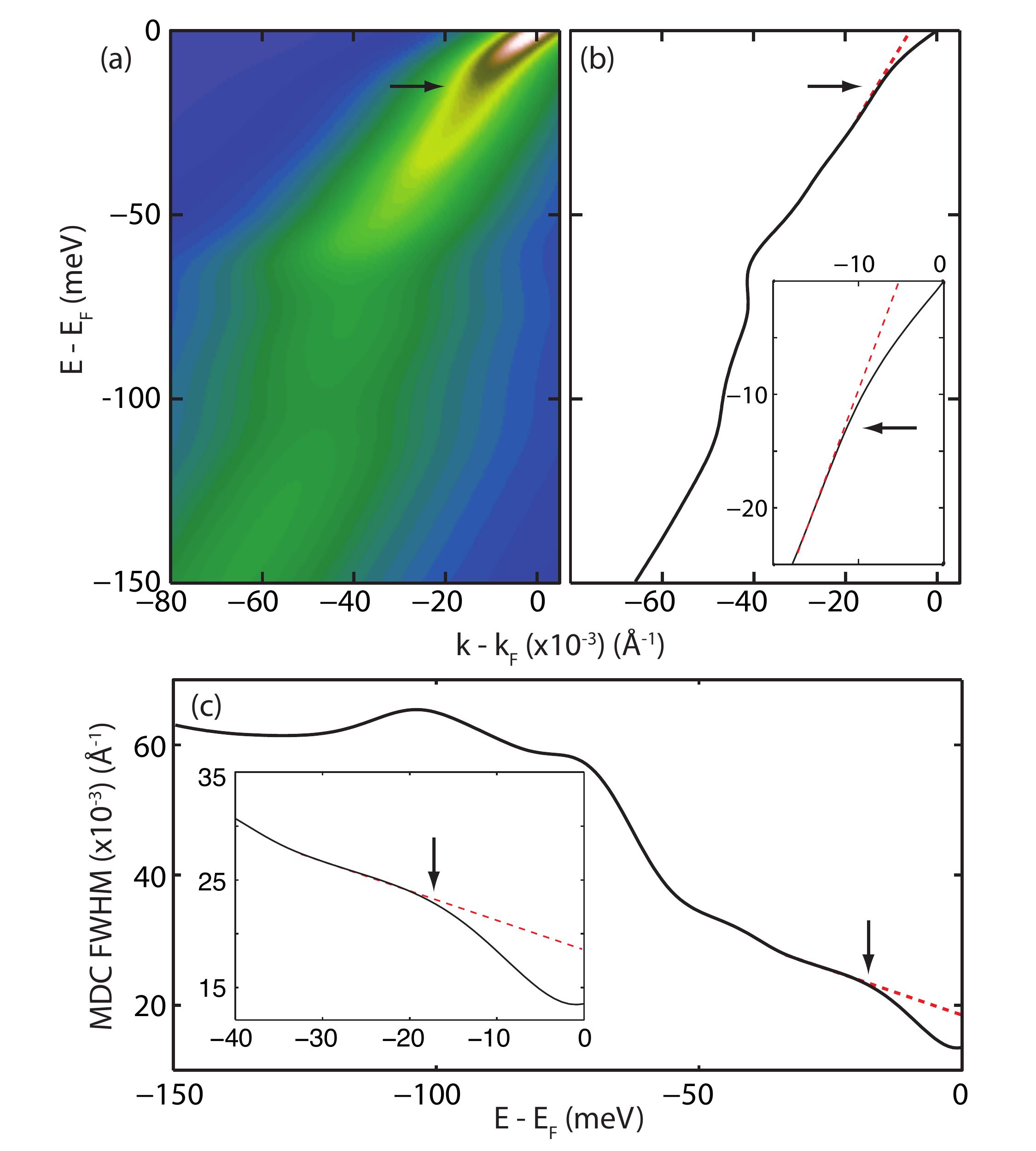}
 \vskip -0.25cm
 \caption{\label{Fig:NodalTheory} (Color online) (a) The calculated spectral 
 function $A(\bk,\omega)$ along the nodal direction $(0,0)$-$(\pi,\pi)$ of superconducting 
 (25 K) Bi-2212. (b) 
 The MDC-derived dispersion and (c) MDC FWHM obtained from the calculated $A(\bk,\omega)$. 
 In calculating $A(\bk,\omega)$, an additional 
 component $\propto\omega^2$ has been added to the imaginary part 
 of the self-energy in order to simulate additional broadening due to 
 el-el interactions.}
\end{figure}

The strength of the el-ph coupling can be estimated from the slope of the 
self-energy at the Fermi level
\begin{equation}
\lambda(\bk) = -\frac{\partial\mathrm{Re}\Sigma(\bk,\omega)}
{\partial \omega}\bigg|_{\omega = 0}.
\end{equation}
For the above choice of parameters we obtain $\lambda(\bk_{node}) = 0.56$ for 
the acoustic branch, consistent with the observed magnitude of the 
renormalization.\cite{VishikPRL2010, Anzai}   
For the optical modes we set the overall strength such that 
$\lambda_{z} = 0.23$, $0.05$ and $0.05$ for the $36$, $55$ and $70$ meV modes, 
respectively, in order to match previous work.\cite{tpdPRL2004,JohnstonPRB2010}  

The calculated $A(\bk,\omega)$ (and corresponding MDC analysis) for nodal cut in 
the superconducting state (25 K) is shown in Fig. 
\ref{Fig:NodalTheory}. As with the experimental data (Fig. \ref{Fig:NodalExp}),  
the low energy kink is evident in the raw spectral function appearing at an energy set 
by the phonon dispersion weighted by $g(\bq)$. 
Since the coupling for scattering to the antinode is suppressed, this energy scale is 
slightly smaller than the van Hove points in the phonon density of states (10.6 and 15 meV).  
The low-energy kink is also evident in the MDC-derived 
dispersion and linewidth, shown in Figs. \ref{Fig:NodalTheory}b and \ref{Fig:NodalTheory}c, 
respectively.  Both are in excellent agreement with the experimental data (Fig. \ref{Fig:NodalExp}), 
demonstrating that the forward scattering nature of the coupling to 
the acoustic branch accounts for both the energy scale of the low-energy kink  
and the overall strength of the renormalization near the Fermi level.   

\begin{figure}[tl]
 \begin{center}
 \includegraphics[width=\columnwidth]{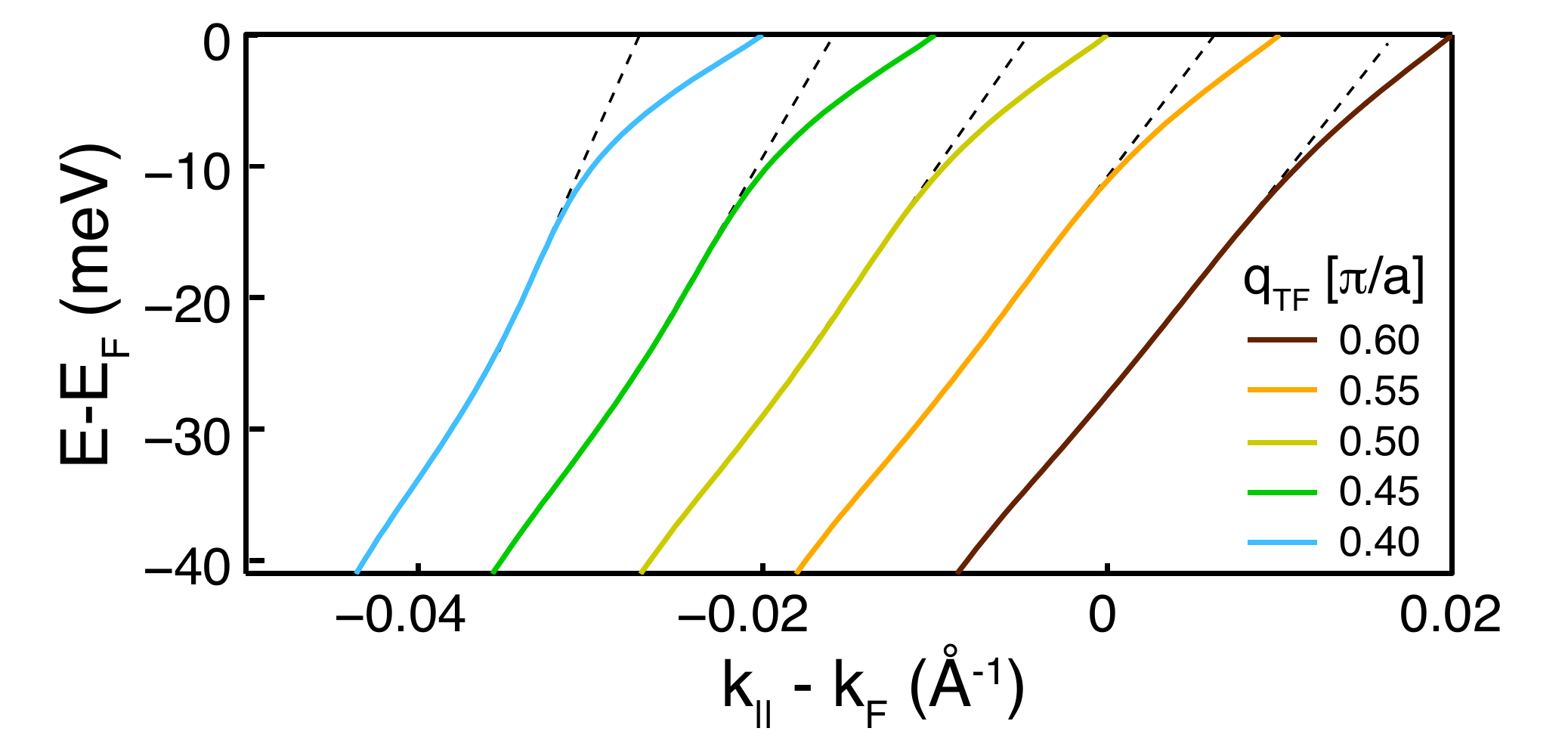}
 \end{center}
 \vskip -0.75cm
 \caption{\label{Fig:Lambda}
 Nodal MDC derived dispersions obtained from the model $A(\bk,\omega)$ for various 
 values of the Thomas-Fermi wavevector. The value of the dielectric constant $\epsilon$ 
 has been held fixed for each case. 
 }
\end{figure}

Since the coupling to the acoustic mode arises from the modulation of the screened 
Coulomb interaction a doping dependence naturally emerges from the increased 
metallicity of the cuprates with overdoping.\cite{JohnstonPRB2010} 
To examine the implications of this, in Fig. \ref{Fig:Lambda} we plot nodal MDC-derived 
dispersions obtained for our model calculations 
while varying $q_{TF}$ from $0.4\pi/a$ to $0.6\pi/a$.  This allows us to mimic the 
increased screening which occurs with progressive overdoping.  For decreasing $q_{TF}$ 
the overall strength of the coupling increases and becomes increasingly peaked at 
small $\bq$. As a result, stronger kinks are produced with 
underdoping in agreement with experiment.\cite{VishikPRL2010}  
We also note that increased coupling at small $\bq$ with underdoping has also 
been reported for more a complicated form for the dielectric function 
based on RPA treatments\cite{JohnstonPRB2010} and is therefore expected to be a 
generic result.  These effects are also expected to survive the inclusion of 
the short-range Hubbard interaction, which largely suppresses the el-ph vertex 
at large $\bq$ over its value at small $\bq$.\cite{Hubbard} 

\begin{figure*} [t]
\includegraphics [width=\textwidth]{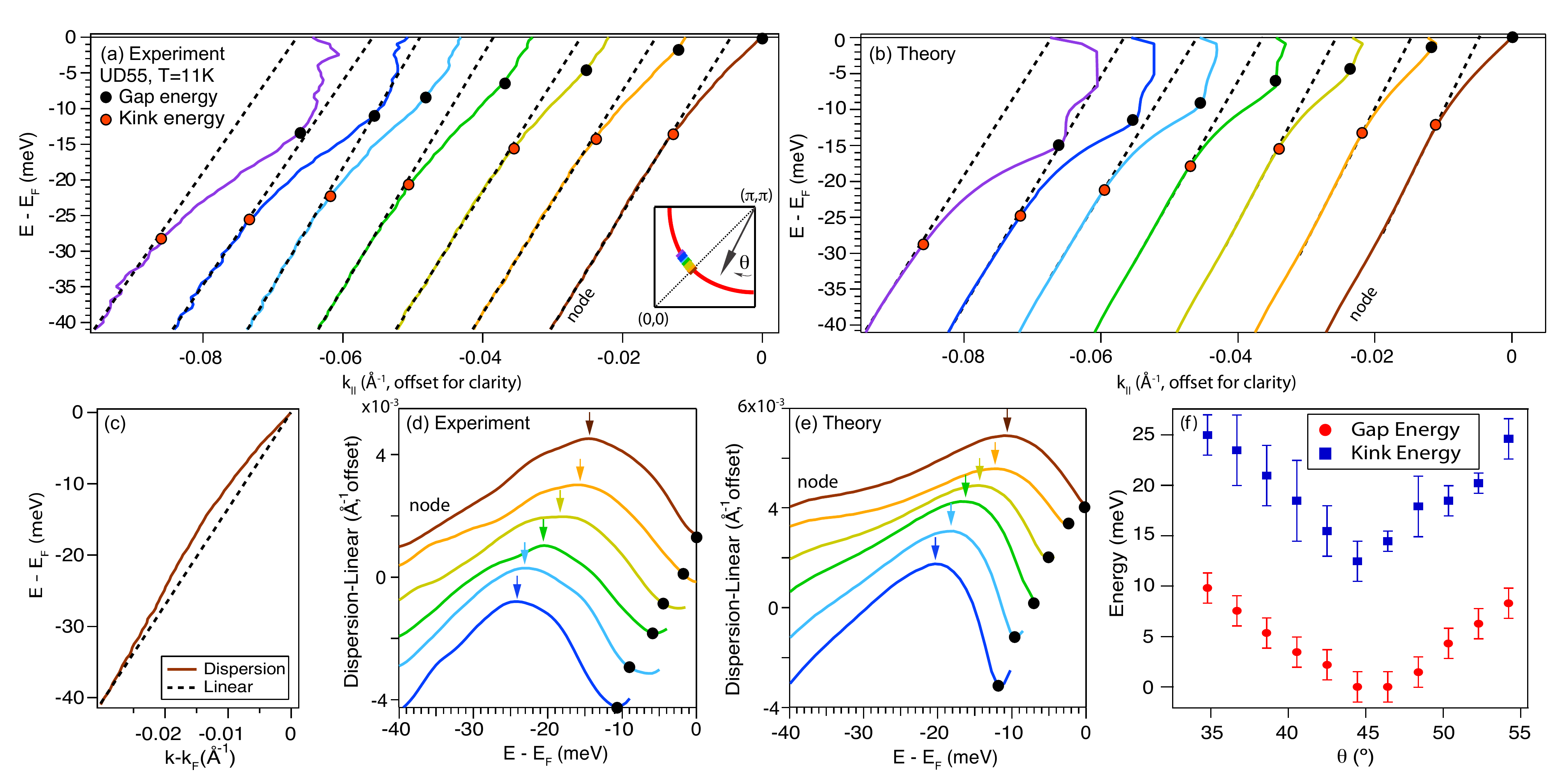}
\centering
\vskip -0.25cm
\caption{\label{Fig:Momentum} 
(Color online) Low energy kink away from node (UD55, 11K).  
(a) MDC dispersions at node (rightmost) and away from the node, offset 
horizontally for clarity.  Black circle represents gap energy, which is 
determined by fitting symmetrized data to a minimal model.
\cite{NormanPRB1998}  Red circle defines the approximate 
low-energy kink position, as determined by where the measured dispersion 
deviates from v$_{mid}$.  (b) MDC dispersions obtained from the 
calculated $A(\bk,\omega)$.
(c) A sketch of how the kink position can be further quantified by 
subtracting a linear offset from the dispersion (shown here for nodal cut) 
between the gap energy and 40 meV. 
(d) Difference in \textbf{k} between dispersion and 
dotted line in (c) for cuts from the node (top) and moving away from the node 
(towards the bottom).  The data has is smoothed over 20 iterations and 
the curves have been offset vertically for clarity.  
Black dots indicate the gap energies and arrows mark peaks of these curves, 
determined from derivative after smoothing. 
(e) The difference in \textbf{k} between the MDC dispersion and linear 
dispersion as in (d) but obtained from the calculated $A(\bk,\omega)$.  
(f) Momentum dependence of gap and kink energies, the 
latter determined by method in (c).}
\vskip -1.0cm
\end{figure*}

We now turn our attention to the momentum dependence of the low-energy kink. 
As previously mentioned, for small $q_{TF}$ the el-ph self-energy is dominated by 
scattering to nearby momentum states and thus samples the local $\bk$-space value of the superconducting 
gap.  Therefore, for off-nodal cuts, the energy of the kink should begin to gap shift, following 
the local value of $\Delta(\bk)$ for each cut.  In Figs. \ref{Fig:Momentum}a and \ref{Fig:Momentum}b 
we present experimental and calculated MDC-derived dispersions for a series of off-nodal cuts 
in a direction parallel to (0,0)-($\pi,\pi$) with cuts taken up to 11 degrees 
away from the node.  
Experimentally, the superconducting gap is determined independently 
for each cut by fitting symmetrized energy distribution curves at $k_F$ to a minimal model 
proposed by Norman {\it et al}..\cite{NormanPRB1998} In Fig. \ref{Fig:Momentum}a the kink position is determined 
from the deviation of the a straight line determined from fitting the 
dispersion slope $v_{mid}$ between 30-40 meV. 
The parameters for the calculations (Fig. \ref{Fig:Momentum}b) are identical to those used in Fig. \ref{Fig:NodalTheory}.

As can be seen in the experimental data, as 
cuts are taken progressively further from the node both the superconducting gap
and kink position shift to higher energies. 
This behavior is  
also reflected in the linear bare band subtraction, shown in Figs. \ref{Fig:Momentum}d  
and is qualitatively reproduced by the model calculations.  
The momentum dependence of both energy scales are summarized in Fig. \ref{Fig:Momentum}f, which 
shows a clear, constant offset between the local value of the gap and the kink position, 
similar to a previous study.\cite{RameauPRB2009} 
Fig. \ref{Fig:Momentum}b shows the calculated MDC dispersions 
and Fig. \ref{Fig:Momentum}e show the subtracted linear bare band (Fig. \ref{Fig:Momentum}c)  
obtained from applying the same procedures to the calculated spectral functions.  
The agreement between the theory and experiment is excellent and 
show that a strong coupling to the in-plane acoustic branch 
also captures the momentum-dependence of the low-energy kink.   

Our results indicate that the acoustic modes couple to the 
lattice in a strongly doping and momentum-dependent way, which have a 
number of interesting implications and general consequences for these and other phonon modes. 
As the Coulomb interaction may 
dramatically alter the momentum-dependence of lattice effects, which 
become more pronounced as the screening length increases, the effect on 
transport can be substantial.\cite{JohnstonPRB2010} Due to the large forward scattering peak, 
coupling to the acoustic branch may become less prominent in transport 
measurements apart from its overall renormalization of the Fermi velocity, 
as noted in a previous study (Ref. \onlinecite{VishikPRL2010}). 
Importantly, the $d_{x^2-y^2}$ pairing interaction from phonons can be dramatically 
changed whereby the screened e-ph interaction which favors large momentum scattering  
can become more peaked at small $\bq$ as screening breaks down and therefore might not as 
detrimental to $d$-wave pairing. This implies that the acoustic modes may 
in fact help $d$-wave pairing, in conjunction with electrons' coupling 
to c-axis and bond-stretching modes, in a way that would not be captured 
in fully itinerant and weak coupling scenarios. This further highlights 
the need to consider lattice effects and coupling to the full spectrum of 
oxygen phonon modes in order to obtain a complete understanding of the 
the physics of the high-Tc cuprates.


The authors thank B. Moritz, R. He, E. van Heumen, 
N. C. Plumb and D. Dessau for useful discussions. 
This work was supported by the US Department of  
Energy, Office of Basic Energy Sciences under contract No. DE-AC02-76SF00515 
and DE-FG03-01ER45929-A001. 
S. J. would like to acknowledge financial support from NSERC and the Foundation 
for Fundamental Research on Matter.  Computational resources were provided by 
the Shared Hierarchical Academic Computing Network.

\end{document}